\documentclass[12pt]{article}
\usepackage{graphicx}
\usepackage{epsfig}
\usepackage{rotating}
\usepackage{dcolumn}
\usepackage{bm}
\newcommand{\comment}[1]{} 
\def \non{\nonumber}
\def \ra{\rightarrow}
\def \bea{\begin{eqnarray}}
\def \eea{\end{eqnarray}}
\def \Qbar{\overline Q}
\def \qqbar{Q\Qbar}
\def \slj{^{2S+1}L_J}

\def \ccbar{c\overline{c}}

\def \sg{\sigma}
\begin{document}
\title{
\vspace{-1.2cm}
\begin{flushright}\vbox{\begin{tabular}{c}
           {\small TIFR/TH/10-21}\\[1.0cm]
\end{tabular}}\end{flushright}
{\bf $\eta_c$ production at the Large Hadron Collider}}
\author{
   Sudhansu~S.~Biswal\footnote{E-mail: sudhansu@theory.tifr.res.in} 
     ~and  K.~Sridhar\footnote{E-mail: sridhar@theory.tifr.res.in} \\[0.2cm]
    {\it \small Department of Theoretical Physics} \\
    {\it \small Tata Institute of Fundamental Research}\\
    {\it \small Homi Bhabha Road, Mumbai 400 005, INDIA}
}
\date{}
\maketitle
\begin{abstract}
\noindent We have studied the production of the $^1S_0$ charmonium state,  
$\eta_c$, at the Large Hadron Collider (LHC) in the 
framework of Non-Relativistic Quantum Chromodynamics (NRQCD) using 
heavy-quark symmetry. We find that NRQCD predicts a large 
production cross-section for this resonance at the LHC even after 
taking account the small branching ratio of $\eta_c$ into two photons. 
We show that it will be possible to test NRQCD through its 
predictions for $\eta_c$, with the statistics that will be achieved 
at the early stage of the LHC, running at a center of mass energy 
of 7 TeV with an integrated luminosity of 100 pb$^{-1}$.   
\end{abstract}

\noindent Non-Relativistic Quantum Chromodynamics (NRQCD)~\cite{bbl} is an 
effective theory that has been extensively used to study
the production and decay of quarkonia. 
NRQCD is derived from the QCD Lagrangian by neglecting all states 
of momenta much larger than the heavy quark mass, $M_Q$ and 
to account for this exclusion by adding new interaction terms 
in the effective Lagrangian. It is then possible to expand the
quarkonium state in-terms of $v$, the relative velocity 
of the heavy quarks in the bound state. In this expansion, 
the $\qqbar$ pair in the intermediate state can be in either
colour-singlet or colour-octet configurations denoted 
by $\qqbar [\slj^{[1,8]}]$. However, the  
colour-octet $\qqbar$ state evolves non-perturbatively into a physical
colour-singlet state by emission of one or more soft gluons.   
The cross section for production of a quarkonium state $H$ can be 
factorised as:
\bea
  \sigma(H)\;=\;\sum_{n=\{\alpha,S,L,J\}} {F_n\over {M_Q}^{d_n-4}}
       \langle{\cal O}^H_n({}^{2S+1}L_J)\rangle, 
\label{factorizn}
\eea
where $F_n$'s are the short-distance coefficients and ${\cal O}_n$ are 
operators of naive dimension $d_n$, describing the long-distance effects.  
These non-perturbative matrix elements are guaranteed to be
energy-independent due to the NRQCD factorization formula, so that they
may be extracted at a given energy and used to predict quarkonium cross-sections
at other energies.

Before the effective theory approach of NRQCD was developed, the 
colour-singlet model (CSM)~\cite{berjon,br} 
was used to analyze the production of quarkonia, where the $Q \bar Q$
state produced in the short-distance process was assumed to be
a colour-singlet. However, it was pointed out in Ref.~\cite{jpsi} that 
contributions from the colour-octet operators are 
significant in describing the phenomenology of large-$p_{_T}$ $P$-state 
charmonium production at the Tevatron \cite{cdf}. 
In Refs. \cite{cho1,cho2} the complete set of short-distance 
coefficients in NRQCD needed to study $J/\psi$ and $\chi$ production
was calculated and compared with the data from Tevatron \footnote{See also 
Ref. \cite{cgmp}. For a detailed review of quarkonium production
see Ref. \cite{brambilla}}. These NRQCD calculations gave a good description
of the shapes of the $p_{_T}$ distributions of the charmonium resonances
at the Tevatron but the normalization of
these distributions was not predicted in NRQCD i.e. the non-perturbative
matrix elements which determined the normalization had to be obtained
by a fit to the data. Independent tests of the effective theory approach
were, therefore, necessary to determine the validity of the approach
and, indeed, various proposals were made \cite{tests}
to test NRQCD. But several of these 
proposals are not for large-$p_{_T}$ quarkonium production and
the validity of NRQCD factorization at low-$p_{_T}$ is suspect.

One interesting test of NRQCD comes from the study of the polarization of
$J/\psi$'s at large-$p_{_T}$ \cite{polar1} which 
primarily comes from a fragmentation-like processes where a single gluon splits
into a $Q \bar Q$ pair which inherits the transverse polarization
of the gluon. The heavy-quark symmetry of NRQCD then comes into play in
protecting this transverse polarization in the non-perturbative evolution
of the $Q \bar Q$ pair into a $J/\psi$. The large-$p_{_T}$ $J/\psi$ is, therefore,
strongly transversely polarized. This is not true at even moderately low $p_{_T}$
where the $J/\psi$ is essentially unpolarized. The $p_{_T}$ dependence of the
polarization is, therefore, a very good test of the theory \cite{polar2}.

The CDF experiment has measured the $p_{_T}$-dependence of the polarization
and they find no evidence for any transverse polarization at large
$p_{_T}$ \cite{polar3}, which seems to indicate a dramatic failure of the theory. 
Inspite of the successful prediction of the production cross-sections of 
the various charmonium resonances it may well be that the effective theory is 
missing out on some aspect of the physics of quarkonium formation. 
Alternatively, it could 
be that the charm quark is too light to be treated in NRQCD. 
On the other hand, 
polarization measurements are usually fraught with problems and it
may well be that the problem is elsewhere. Finally, because the
colour-singlet channel predicts unpolarized $J/\psi$'s, there have
been attempts to increase up the colour-singlet contribution to the
production processes by invoking Reggeized gluons \cite{stirling} 
or enhanced effects of higher-order
QCD corrections in the singlet channel \cite{gong, artoisenet}. 
For reviews of the current status of these
calculations and their experimental consequences, see Refs.~\cite{lansberg1,
lansberg2}.

In this situation, it is worthwhile looking for other tests of NRQCD which
successfully navigate between low-$p_{_T}$ and polarization. The heavy-quark
symmetry of NRQCD provides a set of relations between non-perturbative
parameters of different resonances so a measurement of a given
state yields information on the non-perturbative parameter of
another state related to the former by heavy-quark symmetry. This
fact has been exploited to study $h_c$ production at the Tevatron
\cite{Sridhar:1996vd} and, more recently, at the LHC \cite{Sridhar:2008sc}. 
Similarly $\eta_c$ production
at the Tevatron has also been studied \cite{Mathews:1998nk}. In this paper, we
study the production of $\eta_c$ at the LHC. 

The Fock space expansion of the physical $\eta_c$, which is a $^1S_0$
($J^{PC}=0^{-+}$) state, is: 
\bea
\left|\eta_c\right>={\cal O}(1)
	\,\left|\qqbar[^1S_0^{[1]}] \right>+
	 {\cal O}(v^2)\,\left|\qqbar[^1P_1^{[8]}]\,g \right>+
	{\cal O}(v^4)\,\left|\qqbar[^3S_1^{[8]}]\,g \right>+\cdots ~.
\label{fockexpn}
\eea
In the above expansion the colour-singlet $^1S_0$ state contributes 
at ${\cal O}(1)$. As the $P$-state production is itself down 
by factor of ${\cal O}(v^2)$ both the colour-octet $^1P_1$ and $^3S_1$ 
channels effectively contribute at the same order. The  
colour-octet state $^1P_1^{[8]}$ ($^3S_1^{[8]}$) becomes a 
physical $\eta_c$ by  emitting a gluon in an E1 (M1) transition.  
Keeping terms up-to ${\cal O} (\alpha_s^3 v^7)$ the $\eta_c$ production 
cross section can be parameterized as:
\bea
\sg(\eta_c)&=&
	\frac{F_1[^1S_0]}{M^2}\, \left< 0 \right| {\cal O}_1^{\eta_c}
		[^1S_0]\left| 0 \right> \nonumber\\
     && + \frac{F_8[^1P_1]}{M^4}\, \left< 0 \right| {\cal O}_8^{\eta_c}
		[^1P_1]\left| 0 \right> 
        + \frac{F_8[^3S_1]}{M^2}\, \left< 0 \right| {\cal O}_8^{\eta_c}
		[^3S_1]\left| 0 \right>, 
\eea
where the coefficients, $F$'s, are the cross sections for the production of
$\ccbar$ pair in the respective angular momentum and colour states. 
The differential cross section for $\ccbar$ pair production with 
specific angular momentum and colour states at the LHC is given by:
\bea
&&\frac{d\sg}{dp_{_T}} \;(p p \ra \ccbar\; [^{2S+1}L^{[1,8]}_J]\, X)= \non \\
&&\sum \int \!dy \int \! dx_1 ~x_1\:G_{a/p} (x_1)~x_2\:G_{b/p}(x_2) 
\:\frac{4p_{_T}}{2x_1-\overline{x}_T\:e^y}\:\frac{d\hat{\sg}}{d\hat{t}}
(ab\ra \ccbar[^{2S+1}L_J^{[1,8]}]\;d),
\label{eq:diff}
\eea
where the summation is over the partons ($a$ and $b$),    
the final state $\qqbar$ is in the $^1 S^{[1]}_0$, $^1 P^{[8]}_1$,  
$^3 S^{[8]}_1$ states and $G_{a/p}$,   
$G_{b/p}$ are the distributions of partons $a$ and $b$ in the
protons and $x_1$, $x_2$ are the respective  momentum they carry.
$x_2$ is related with $x_1$ as: 
\bea
x_2=\frac{x_1\:\overline{x}_T\:e^{-y}-2\tau}{2x_1-\overline{x}_T\:e^{y}},
\eea
where $\overline{x}_T=\sqrt{x_T^2+4\tau} \equiv 2 M_T/\sqrt{s}$ \ with \  
$x_T=2p_{_T}/\sqrt{s}$ and  \(\tau=M^2/s\).
Here $\sqrt{s}$ is the center-of-mass energy, 
$M$ is the mass of the resonance and $y$ is the rapidity at which the 
resonance is produced. 
The subprocesses contributing to Eq.(\ref{eq:diff}) are:
\bea
g ~g &\ra& \qqbar[\slj] ~g,\non \\
g~ q (\bar q)  &\ra& \qqbar[\slj] ~q (\bar q),\\ 
q ~\bar q &\ra& \qqbar[\slj] ~g. \non
\eea
The matrix  elements for the subprocesses 
corresponding to $F_1[^1S_0]$ and $F_8[^3S_1]$ are listed in 
Refs.~\cite{cho2,gtw}. The remaining coefficient $F_8[^1P_1]$ has been 
calculated and used in~\cite{Mathews:1998nk}  
to analyze $\eta_c$ production at the Tevatron. 

We use heavy quark spin-symmetry to obtain the values of  
$\left< {\cal O}_n^{\eta_c}\right>$'s from the experimentally 
predicted values $\langle {\cal O}_n^{J/\psi}\rangle$'s. Using this 
symmetry the $\langle {\cal O}_n^H \rangle$'s are related as:
\bea
\left< 0 \right|{\cal O}_1^{\eta_c}[^1S_0]\left| 0\right>&=&
{1 \over 3} \left< 0 \right|{\cal O}_1^{J/\psi}[^3S_1]\left| 0 \right>
\;(1+ O(v^2)), \non\\
\left< 0 \right|{\cal O}_8^{\eta_c}[^1P_1]\left| 0\right>&=&
\left< 0 \right|{\cal O}_8^{J/\psi}[^3P_0]\left| 0 \right>
\;(1+ O(v^2)), \non\\
\left< 0 \right|{\cal O}_8^{\eta_c}[^3S_1]\left| 0\right>&=&
\left< 0 \right|{\cal O}_8^{J/\psi}[^1S_0]\left| 0 \right>
\;(1+ O(v^2)).
\label{Ovalues}
\eea
We use the predicted value of the singlet matrix elements listed in 
Ref.~\cite{cho1} 
( ${\cal C}_1 \equiv$ $\left< 0 \right|{\cal O}_1^{J/\psi}[^3S_1]
\left| 0 \right>=1.2$ GeV$^3$) and 
the octet matrix elements extracted from the 
CDF $J/\psi$ data~\cite{cho2}  (i. e. ${\cal C}_2+{\cal C}_3 \equiv$ 
\(\frac{\left< 0 \right|{\cal O}_8^{J/\psi}[^3P_0]\left| 0 \right>}
{M_c^2}+
\frac{\left< 0 \right|{\cal O}_8^{J/\psi}[^1S_0]\left| 0 \right>}{3}
=(2.2\pm0.5)\times 10^{-2}\) GeV$^3$). It is to be noted that only a linear 
combination ${\cal C}_2~ +~ {\cal C}_3$ can be extracted from the CDF data as 
the shapes of ${\cal C}_2$ and ${\cal C}_3$ contributions to the $J/\psi$ 
$p_{_T}$-distribution are almost identical. Hence, for predicting
the $\eta_c$ rate
we assume that either ${\cal C}_2$ or ${\cal C}_3$ saturates the sum
so that our predictions indicate the band within which we expect the
experimental value of the $\eta_c$ production cross-section to lie. 
More explicitly, we consider the two extreme cases where in the first case, 
the maximum possible contribution 
is from the $^3S^{[8]}_1$ channel and none from the $^1P^{[8]}_1$ 
channel whereas in the second case the $^1P^{[8]}_1$ contributes its maximum 
while the $^3S^{[8]}_1$ channel does not contribute. 

%
\begin{figure}[!h]
\centerline{
\epsfig{file=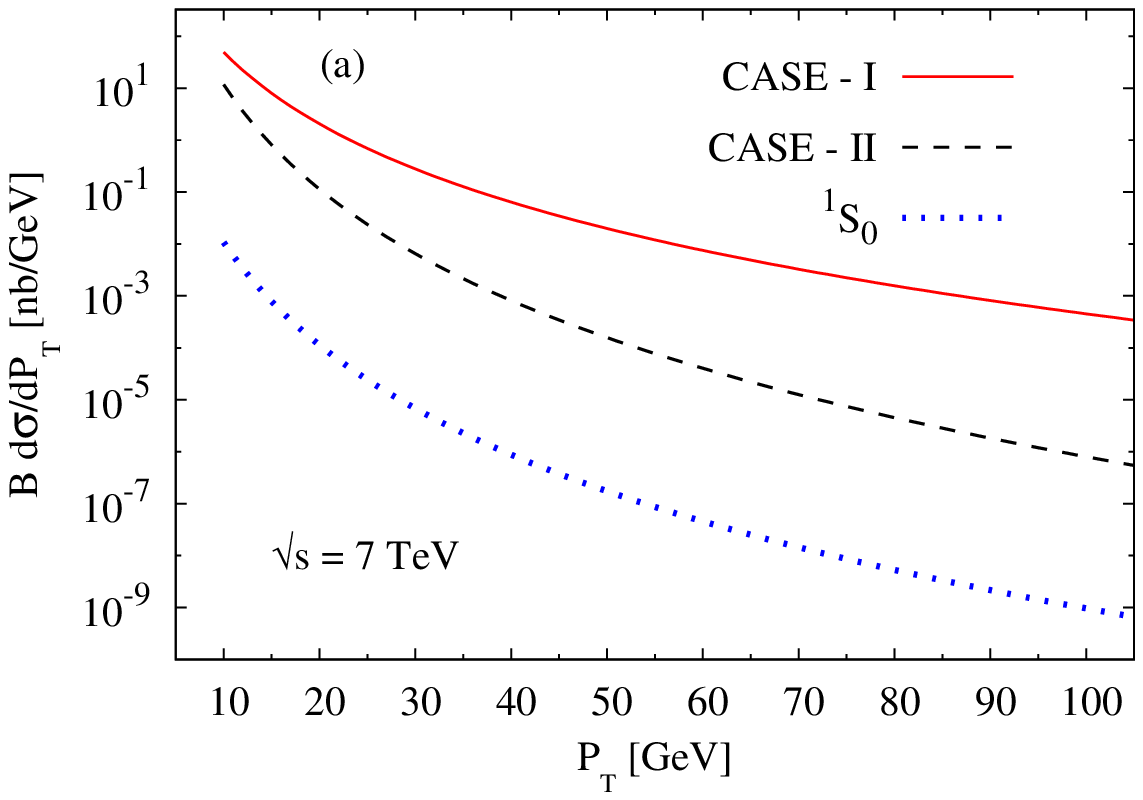, width=8.0cm, totalheight=7.5cm}
\epsfig{file=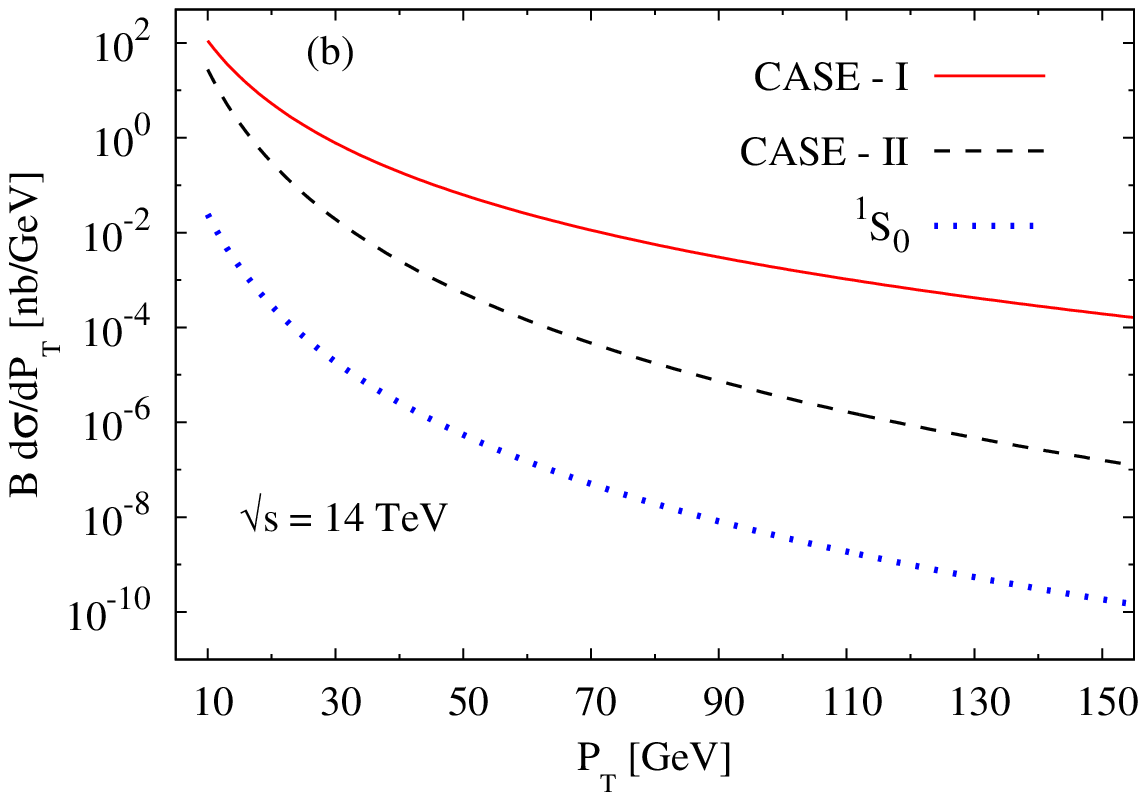, width=8.0cm, totalheight=7.5cm}}
\caption{\label{fig-1} $d\sigma/dp_{_T}$ (in nb/GeV) for $\eta_c$ production (after folding in
with Br($\eta_c \rightarrow \gamma\gamma)=3.0\times 10^{-4}$) in
$p  p$ collisions at $\sqrt{s} =$ 7~TeV and 14~TeV with $-2 \le y \le 2$. }
\end{figure}
In Fig.~\ref{fig-1} 
we have displayed the differential cross section $Bd\sg/dp_{_T}$ 
as a function of $p_{_T}$ at two different center-of-mass energies, 
viz. $\sqrt{s} =$ 7 and 14 TeV respectively, where $B$
is the $\eta_c \ra \gamma \gamma$ branching ratio ($B=3 \times 10^{-4}$). 
We have used CTEQ 5L LO parton densities \cite{cteq} evolved to a scale 
$Q=M_T$. In both (a) and (b) of Fig. 1, we have curves marked I and II
where I is the sum of the colour-singlet and the $^3S_1^{[8]}$ contributions
and II is, likewise, the sum of the colour-singlet and the
$^1P_1^{[8]}$ contribution. In addition, we also display the colour-singlet
curve in both the figures, to bring out the fact that
the octet contributions overwhelmingly dominate the
cross-section. 

To account for the experimental threshold in $p_{_T}$ for the photons
which is about 10 GeV, we use a lower-$p_{_T}$ cut of 20 GeV
on the $\eta_c$ to calculate the integrated cross-sections.
For the LHC running at $\sqrt{s}=7$ TeV, with an integrated 
luminosity ($\cal L$) of 100 pb$^{-1}$, we find that the number of events in 
the $\gamma \gamma$ channel from the singlet $^1S_0^{[1]}$ is about 
40 while the number of events from $^3S_1^{[8]}$ is about 
$10^6$  when $^1P_1^{[8]}$ contribution is neglected, 
while the number of events from $^1P_1^{[8]}$ is about 37660  when 
$^3S_1^{[8]}$  contribution is absent. For the case of 
$\sqrt{s}=14$ TeV and $\cal L =$ 100 pb$^{-1}$, the respective numbers 
would be 100, $2.8 \times 10^6$ and $10^5$.  
Thus the minimum $\eta_c \ra \gamma \gamma$ events at the LHC 
running at $\sqrt{s}=7$ and 14 TeV will be 37700 and $10^5$ 
respectively. 
   
We can see from Fig.~\ref{fig-1} that the shapes of the $^3S_1^{[8]}$ and 
the $^1P_1^{[8]}$ contributions to the $p_{_T}$-distribution is different
and may allow the non-perturbative matrix elements  ${\cal C}_2$ and 
${\cal C}_3$ to be individually determined. We have noted earlier that
the $J/\psi$ cross-section does not provide such a separation. But it is
also pertinent to note that the integrated cross-sections are also very
sensitive to the the values of  ${\cal C}_2$ and ${\cal C}_3$ and
so the measurement of the integrated cross-section alone may provide
this discriminatory ability.
%
\begin{figure}[!h]
\centerline{
\epsfig{file=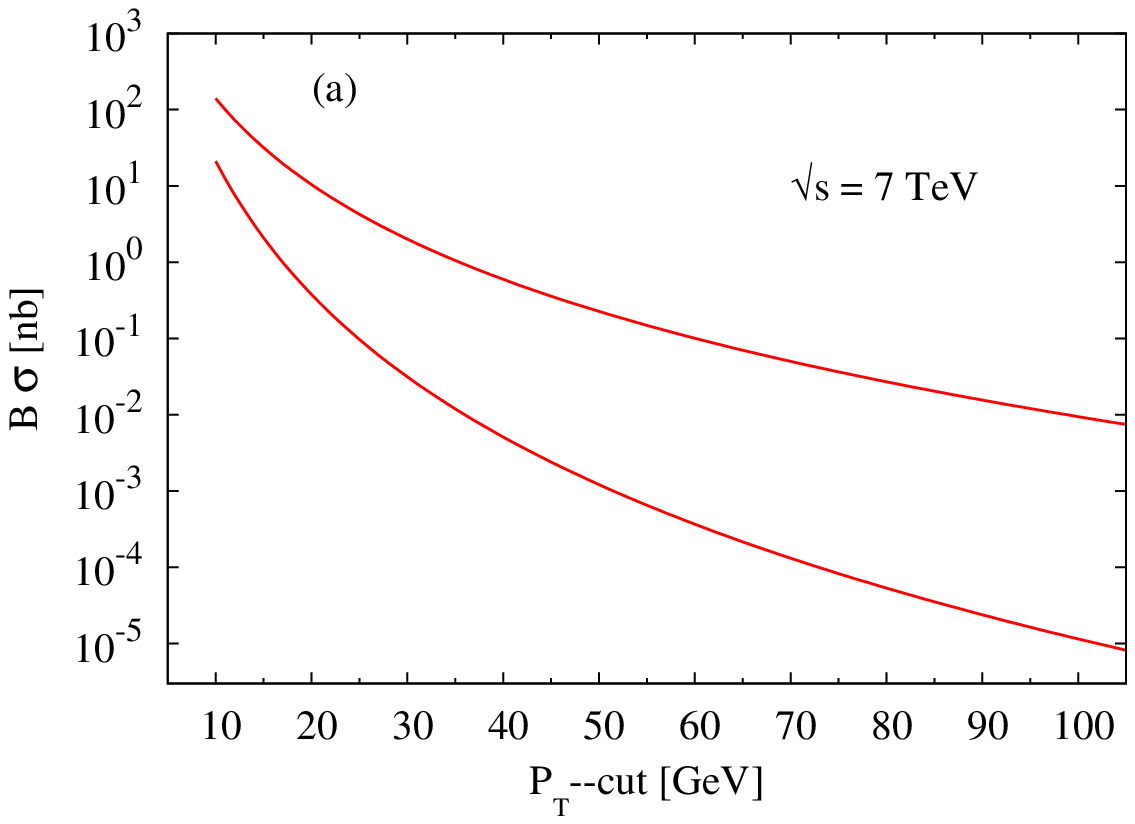, width=8.0cm, totalheight=7.5cm}
\epsfig{file=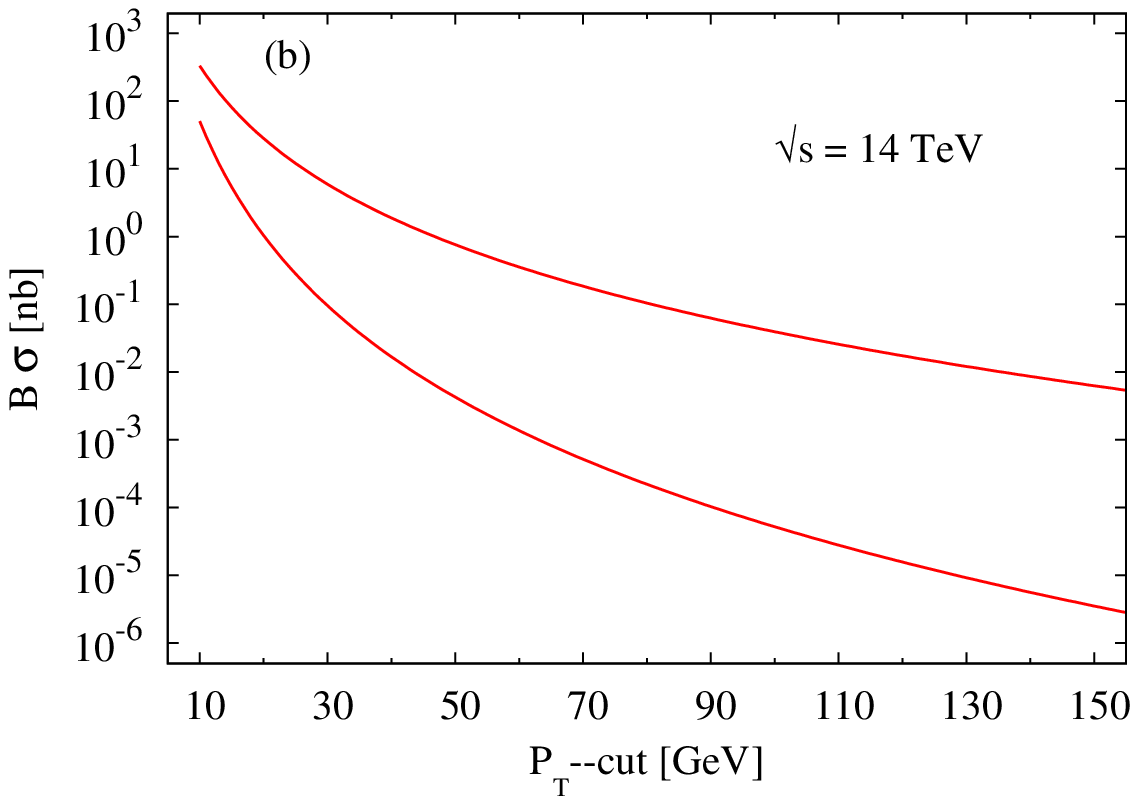, width=8.0cm, totalheight=7.5cm}}
\caption{\label{fig-2} Variation of total
cross section with respect to chosen minimum $p_{_T}$-cut  
for $\eta_c$ production (after folding in
with Br($\eta_c \rightarrow \gamma\gamma)=3.0\times 10^{-4}$) in
$p  p$ collisions at $\sqrt{s} =$ 7~TeV and 14~TeV with $-2 \le y \le 2$. }
\end{figure}

In Fig.~\ref{fig-2} 
we have displayed the effect of increasing the $p_{_T}$-cut on the
magnitude of the integrated cross-section. As explained earlier,
the cut on $p_{_T}$ of the $\eta_c$ will be determined by the minimum
$p_{_T}$ threshold that the experiments use to trigger on the photons, for which
the usual choice is 10 GeV. In case the experiments use a larger
cut on the $p_{_T}$ of the photons in order to improve the quality of
their signal, the $p_{_T}$ cut on the $\eta_c$ will be correspondingly
higher. We see, from Fig. 2, that even when the
cut on the minimum $p_{_T}$ is as large as 50 GeV, the cross-section is
substantial. 
\begin{table}[!h]
\begin{center}
\begin{tabular}{||c|ccc||}
\cline{2-4}
\multicolumn{1}{c|}{} & \multicolumn{3}{c||}
{$\sim$ \rm Number of $\eta_c$ events evaluated using CTEQ 5L LO densities} \\
\hline \hline
\multicolumn{1}{||c|}{}&&&\\
\multicolumn{1}{||c|}{$\sqrt{s}$}
 & $Q = M_T/2$ & $Q = M_T$ & $Q = 2~M_T$  \\[2mm]
\multicolumn{1}{||c|}{}&&&\\
\multicolumn{1}{||c|}{7 TeV} & 
$5.6 \times 10^4$ - $1.5 \times 10^6$& 
$3.8 \times 10^4$ -$10^6$ & $2.6 \times 10^4$ - $7.3 \times 10^5$ 
\\[2mm]
\multicolumn{1}{||c|}{}&&&\\
\multicolumn{1}{||c|}{14 TeV} 
& $1.5 \times 10^5$ - $3.9 \times 10^6$
& $10^5$ - $2.8 \times 10^6$  
& $7.6 \times 10^4$ - $2.1 \times 10^6$
\\[2mm]
\hline \hline
\multicolumn{1}{c|}{} & \multicolumn{3}{c||}
{$\sim$ \rm Number of events evaluated using MRST LO densities} \\
\hline \hline
\multicolumn{1}{||c|}{}&&&\\
\multicolumn{1}{||c|}{$\sqrt{s}$}
 & $Q = M_T/2$ & $Q = M_T$ & $Q = 2~M_T$  \\[2mm]
\multicolumn{1}{||c|}{}&&&\\
\multicolumn{1}{||c|}{7 TeV} 
& $4.9 \times 10^4$ - $1.3 \times 10^6$
& $3.4 \times 10^4$ - $9.5 \times 10^5$   
& $2.4 \times 10^4$ - $6.8 \times 10^5$
\\[2mm]
\multicolumn{1}{||c|}{}&&&\\
\multicolumn{1}{||c|}{14 TeV} 
& $1.2 \times 10^5$ - $3.3 \times 10^6$ 
& $9 \times 10^4$ - $2.4 \times 10^6$ 
& $6.8 \times 10^4$ - $1.8 \times 10^6$ 
\\[2mm]
\hline
\hline
\end{tabular}
\vskip -0.2cm
\caption{\label{tab:events}
Minimum and maximum number of $\eta_c$ events expected at the LHC 
for an integrated luminosity of 100 pb$^{-1}$. 
}
\end{center}
\end{table}
%
 
We also have analyzed the effect on the cross-section of the variation of 
the QCD scale, 
the parton densities and the non-perturbative matrix elements. 
Table~ \ref{tab:events} shows the variation in the minimum and maximum 
number of $\eta_c$ events with scale $Q$, expected at the LHC 
running at $\sqrt{s}=7$ and 14 TeV. 
We find that 
the cross section decreases by about 25-30\% changing the scale from 
$Q=M_T$ to $Q = 2 M_T$ and it increases by 40-50\% for 
the scale choice $Q=M_T/2$ instead of $Q=M_T$. We have checked that 
the cross section decreases by about 10-20\% if we use 
MRST LO densities~\cite{mrs} instead of CTEQ5L LO~\cite{cteq} densities.  
Since the heavy-quark symmetry is an approximate symmetry we 
can expect about 30\% variation in the values of the non-perturbative 
matrix elements we have used. 
All through we have considered only the direct production of 
$\eta_c$ at the LHC. However, an additional contribution to   
$\eta_c$ signal comes from the decays of $J/\psi$. 
We have estimated that this additional contribution to the signal coming 
from $J/\psi$ decays can change our predictions by about 1\% as 
Br($J/\psi \rightarrow  \eta_c \gamma) \sim O(10^{-1})$ and the 
$J/\psi$ production cross section is expected to be of same order  
of $\eta_c$ production cross section.

We would like to remark that such an analysis may also be carried out
for the bottomonium resonance $\eta_b$. The corresponding non-perturbative
parameters in that case, however, are very poorly determined and suffer
from large errors. 

In conclusion, the heavy-quark symmetry of NRQCD allows us to make
predictions for $\eta_c$ production at the LHC.
Measurements of the integrated cross-section and the $p_{_T}$ 
distribution of $\eta_c$ at the LHC will provide a very good test of
NRQCD. 
We show that NRQCD 
predicts a large cross-section for $\eta_c$ at the LHC even
at $\sqrt{s} = 7$ TeV and so this prediction is easily testable. 
%

\section*{Acknowledgments}
We thank Gobinda  Majumder and Gagan Mohanty for useful discussions.  
K.S. completed part of the work presented here on a visit to the 
Department of Physics and Astronomy, University of Southampton, U.K. 
He gratefully acknowledges support from the University of Southampton
for this visit.


\begin{thebibliography}{99}

\bibitem{bbl} G.~T.~Bodwin, E.~Braaten and G.~P.~Lepage,
  Phys.\ Rev.\  D {\bf 51} (1995) 1125
  [Erratum-ibid.\  D {\bf 55} (1997) 5853]
  [arXiv:hep-ph/9407339].
%
%
\bibitem{berjon} E.~L.~Berger and D.~L.~Jones,
  Phys.\ Rev.\  D {\bf 23} (1981) 1521.

\bibitem{br} R.~Baier and R.~Ruckl,
  Z.\ Phys.\  C {\bf 19} (1983) 251.

\bibitem{jpsi} E.~Braaten, M.~A.~Doncheski, S.~Fleming and M.~L.~Mangano,
  Phys.\ Lett.\  B {\bf 333} (1994) 548
  [arXiv:hep-ph/9405407]; D.~P.~Roy and K.~Sridhar,
  Phys.\ Lett.\  B {\bf 339} (1994) 141
  [arXiv:hep-ph/9406386]; M.~Cacciari and M.~Greco,
  Phys.\ Rev.\ Lett.\  {\bf 73} (1994) 1586
  [arXiv:hep-ph/9405241].

\bibitem{cdf} F.~Abe {\it et al.}  [CDF Collaboration],
  Phys.\ Rev.\ Lett.\  {\bf 69} (1992) 3704;
  Phys.\ Rev.\ Lett.\  {\bf 79} (1997) 572; 
  Phys.\ Rev.\ Lett.\  {\bf 79} (1997) 578; D.~E.~Acosta {\it et al.}  [CDF Collaboration],
  Phys.\ Rev.\  D {\bf 71} (2005) 032001
  [arXiv:hep-ex/0412071].
%
\bibitem{cho1} 
P.~L.~Cho and A.~K.~Leibovich,
  Phys.\ Rev.\  D {\bf 53} (1996) 150
  [arXiv:hep-ph/9505329].

\bibitem{cho2} 
P.~L.~Cho and A.~K.~Leibovich,
  Phys.\ Rev.\  D {\bf 53} (1996) 6203
  [arXiv:hep-ph/9511315].

\bibitem{cgmp} M.~Cacciari, M.~Greco, M.~L.~Mangano and A.~Petrelli,
  Phys.\ Lett.\  B {\bf 356} (1995) 553
  [arXiv:hep-ph/9505379].

\bibitem{brambilla}
  N.~Brambilla {\it et al.}  [Quarkonium Working Group],
  arXiv:hep-ph/0412158.

\bibitem{tests} M.~Cacciari and M.~Kramer,
  Phys.\ Rev.\ Lett.\  {\bf 76} (1996) 4128
  [arXiv:hep-ph/9601276]; J.~Amundson, S.~Fleming and I.~Maksymyk,
  Phys.\ Rev.\  D {\bf 56} (1997) 5844
  [arXiv:hep-ph/9601298];
S.~Gupta and K.~Sridhar,
Phys.\ Rev.\  D {\bf 54} (1996) 5545
  [arXiv:hep-ph/9601349]; Phys.\ Rev.\  D {\bf 55} (1997) 2650
  [arXiv:hep-ph/9608433]; M.~Beneke and I.~Z.~Rothstein,
  Phys.\ Rev.\  D {\bf 54} (1996) 2005
  [Erratum-ibid.\  D {\bf 54} (1996) 7082]
  [arXiv:hep-ph/9603400]; W.~K.~Tang and M.~Vanttinen,
  Phys.\ Rev.\  D {\bf 54} (1996) 4349
  [arXiv:hep-ph/9603266];
E.~Braaten and Y.~Q.~Chen,
  Phys.\ Rev.\ Lett.\  {\bf 76} (1996) 730
  [arXiv:hep-ph/9508373];
  K.~M.~Cheung, W.~Y.~Keung and T.~C.~Yuan,
  Phys.\ Rev.\ Lett.\  {\bf 76} (1996) 877
  [arXiv:hep-ph/9509308]; P.~L.~Cho,
  Phys.\ Lett.\  B {\bf 368} (1996) 171
  [arXiv:hep-ph/9509355];
K.~M.~Cheung, W.~Y.~Keung and T.~C.~Yuan,
  Phys.\ Rev.\  D {\bf 54} (1996) 929
  [arXiv:hep-ph/9602423];
P.~Ko, J.~Lee and H.~S.~Song,
  Phys.\ Rev.\  D {\bf 53} (1996) 1409
  [arXiv:hep-ph/9510202];
G.~T.~Bodwin, E.~Braaten, T.~C.~Yuan and G.~P.~Lepage,
  Phys.\ Rev.\  D {\bf 46} (1992) 3703
  [arXiv:hep-ph/9208254].

%
%

\bibitem{polar1} P.~L.~Cho and M.~B.~Wise,
  Phys.\ Lett.\  B {\bf 346} (1995) 129
  [arXiv:hep-ph/9411303].

\bibitem{polar2} M.~Beneke and M.~Kramer,
  Phys.\ Rev.\  D {\bf 55} (1997) 5269
  [arXiv:hep-ph/9611218].

\bibitem{polar3} A.~A.~Affolder {\it et al.}  [CDF Collaboration],
  Phys.\ Rev.\ Lett.\  {\bf 85} (2000) 2886
  [arXiv:hep-ex/0004027]; A.~Abulencia {\it et al.}  [CDF Collaboration],
  Phys.\ Rev.\ Lett.\  {\bf 99} (2007) 132001
  [arXiv:0704.0638 [hep-ex]].



\bibitem{stirling}
  V.~A.~Khoze, A.~D.~Martin, M.~G.~Ryskin and W.~J.~Stirling,
  Eur.\ Phys.\ J.\  C {\bf 39} (2005) 163
  [arXiv:hep-ph/0410020].

\bibitem{gong}
  B.~Gong, X.~Q.~Li and J.~X.~Wang,
  arXiv:0805.4751 [hep-ph].

\bibitem{artoisenet}
  P.~Artoisenet, J.~Campbell, J.~P.~Lansberg, F.~Maltoni and F.~Tramontano,
  Phys.\ Rev.\ Lett.\  {\bf 101} (2008) 152001
  [arXiv:0806.3282 [hep-ph]].

\bibitem{lansberg1}
  J.~P.~Lansberg {\it et al.},
  AIP Conf.\ Proc.\  {\bf 1038} (2008) 15
  [arXiv:0807.3666 [hep-ph]].

\bibitem{lansberg2}
  J.~P.~Lansberg,
  arXiv:0811.4005 [hep-ph].

\bibitem{Sridhar:1996vd}
  K.~Sridhar,
  Phys.\ Rev.\ Lett.\  {\bf 77} (1996) 4880
  [arXiv:hep-ph/9609285].

\bibitem{Sridhar:2008sc}
  K.~Sridhar,
  Phys.\ Lett.\  B {\bf 674} (2009) 36
  [arXiv:0812.0474 [hep-ph]].

\bibitem{Mathews:1998nk}
  P.~Mathews, P.~Poulose and K.~Sridhar,
  Phys.\ Lett.\  B {\bf 438} (1998) 336
  [arXiv:hep-ph/9803424].

%
\bibitem{gtw} R.~Gastmans, W.~Troost and T.~T.~Wu,
  Nucl.\ Phys.\  B {\bf 291} (1987) 731.


\bibitem{cteq}
  H.~L.~Lai {\it et al.}  [CTEQ Collaboration],
  Eur.\ Phys.\ J.\  C {\bf 12}, (2000) 375 
  [arXiv:hep-ph/9903282].
%
\bibitem{mrs} 
  A.~D.~Martin, W.~J.~Stirling and R.~G.~Roberts,
  Phys.\ Lett.\  B {\bf 306} (1993) 145
  [Erratum-ibid.\  B {\bf 309} (1993) 492].
%
\end{thebibliography}
\end{document}